\documentclass[prd,a4paper,tightenlines,eqsecnum,nofootinbib,preprint,floatfix,showpacs]{revtex4}
\usepackage{epsfig}
\usepackage{graphicx}
\usepackage{bm}

\def\bk{{\bm{k}}}
\def\bn{{\bm{n}}}
\def\bN{\bm{N}}

\def\ihat{\hat\imath}\def\Tr{\text{Tr}\,}
\def\tr{\text{tr}\,}

\def\Eq#1{Eq.~(\ref{#1})}

\begin{document}
\title{Anisotropic Goldstone bosons
of strong-coupling lattice QCD at high density}
\author{Barak Bringoltz}
\author{Benjamin Svetitsky}
\affiliation{School of Physics and Astronomy, Raymond and Beverly
Sackler Faculty of Exact Sciences, Tel Aviv University, 69978 Tel
Aviv, Israel}
\begin{abstract}
We calculate the spectrum of excitations in strong-coupling
lattice QCD in a background of fixed baryon density, at a
substantial fraction of the saturation density. We employ a
next-nearest-neighbor fermion formulation that possesses the
$SU(N_f)\times SU(N_f)$ chiral symmetry of the continuum theory.
We find two types of massless excitations: type I Goldstone bosons
with linear dispersion relations and type II Goldstone bosons with
quadratic dispersion relations. Some of the type I bosons
originate as type II bosons of the nearest-neighbor theory.
Bosons of either type can develop anisotropic dispersion relations,
depending on the value of $N_f$ and the baryon density.

\end{abstract}
 \maketitle

\section{Introduction}

In a previous
paper \cite{paper1} we constructed a framework for
calculating the effects of a background baryon density in Hamiltonian lattice
QCD at strong coupling. We used strong
coupling perturbation theory to write an effective Hamiltonian for
color singlet objects \cite{SDQW,Smit}.
At lowest order we obtained an antiferromagnetic
Hamiltonian that describes meson physics with a fixed baryon
background distribution. (Baryons move only at higher order.) The
Hamiltonian was then transformed to the path integral
of a nonlinear $\sigma$ model. The latter is most easily studied
at large $N_c$.

The global symmetry group of the $\sigma$ model depends on the
fermion kernel of the lattice QCD Hamiltonian.
For $N_f$ flavors of naive fermions we get an interaction between
nearest-neighbor (NN) sites that is invariant under $U(N)$ with
\begin{equation}
N=4N_f.
\end{equation}
This symmetry is too large and is indicative of species doubling.
We add next-nearest-neighbor (NNN) interactions to the kernel and reduce
the symmetry to
\begin{equation}
U(N_f)_L \times U(N_f)_R,
\end{equation}
which is almost the symmetry of the continuum theory. The unwanted
$U(1)_A$ is inevitable if one starts with a local, chirally
symmetric theory \cite{NN}. It can easily be broken by hand in the
$\sigma$ model and we ignore it.

In \cite{paper1} we studied the NN theory and found that the
ground state breaks $U(N)$ spontaneously. The breakdown pattern
depends on the baryon density. In \cite{odo} it was shown that the
excitations of the NN theory divide into two types: type I
Goldstone bosons with linear dispersion relations and type II
Goldstone bosons with quadratic dispersion relations. These
excitations fit the pattern described by Nielsen and Chadha
\cite{Nielsen:hm} and studied by
Leutwyler~\cite{Leutwyler:1993gf}. Type II Goldstone bosons,
typical of ferromagnets, are prominent in work on effective field
theories for dense QCD \cite{Schafer,Sannino}.

The NNN interactions may be treated as a perturbation that
removes some of the global degeneracy of the NN vacuum.
In Ref.~\onlinecite{NNN} we found that for all baryon densities studied,
the ground state breaks the NNN theory's axial symmetries.
In all cases with nonzero baryon density, the discrete rotational symmetry is broken as well.
In this paper we investigate how the NNN interactions affect the Goldstone boson
spectrum, completing our picture of the lattice theory that has the symmetry
of continuum QCD.

\section{Non-linear $\sigma$ model}
\label{sec:NLSM}
We give here a brief description of the elements comprising the $\sigma$ model.
More details may be found in \cite{paper1}.

The $\sigma$ field at
site $\bn$ is an $N \times N$ hermitian, unitary matrix given by a $U(N)$
rotation of the reference
matrix $\Lambda$,
\begin{equation}
\sigma_\bn=U_\bn \Lambda U^{\dag}_\bn \label{sigma_n},
\end{equation}
with
\begin{equation}
\Lambda = \left( \begin{array}{cc}
        \bm1_{m}     & 0 \\
        0 &     -\bm1_{N-m}        \end{array}     \right). \label{Lambda}
\end{equation}
The $\sigma$ field thus represents an element of the coset space
$U(N)/[U(m)\times U(N-m)]$.
The number $m$ can vary from site to site and is determined by the local baryon
number $B_\bn$ according to
\begin{equation}
m=B_\bn+N/2.
\end{equation}

The Euclidean action is
\begin{equation}
\label{S} S=\frac{N_c}2\int d\tau\left[-\sum_{\bn} \Tr\Lambda
U^{\dag}_{\bn} \partial_\tau U_{\bn} +\frac {J_1}2\sum_{\bn i} \Tr
\left( \sigma_\bn \sigma_{\bn+\ihat} \right) + \frac{J_2}2
\sum_{\bn i} \Tr \left( \sigma_\bn \alpha_i \sigma_{\bn+2\ihat}
\alpha_i\right)  \right].
\end{equation}
Here $\alpha_i$ is the $4\times4$ Dirac matrix, times the unit matrix
in flavor space.
The NN term is invariant under the global $U(N)$ transformation
$U_\bn \to VU_\bn$ (or $\sigma_\bn \to V \sigma_\bn V^\dag$) while the NNN term
is only
invariant if $V^\dag \alpha_i V=\alpha_i$ for all $i$.  This
restricts $V$ to the form
\begin{equation}
V=\exp \left[ i\left(\theta_V^a+\gamma_5\theta_A^a\right)\lambda^a \right],
\end{equation}
where $\lambda^a$ are flavor generators.  This is
a chiral transformation in $U(N_f)\times U(N_f)$.
[The $U(1)$ corresponding to baryon number is realized trivially on
$\sigma_\bn$.]

The NNN term couples (discrete) spatial rotations to the internal symmetry,
{\em viz.},
\begin{equation}
\sigma_\bn \to R^{\dag} \sigma_{\bn'} R, \qquad \bn'={\cal R} \bn.
\label{rotattions}
\end{equation}
Here ${\cal R}$ is a $90^\circ$ lattice rotation and $R$ represents it
according to
\begin{equation}
R=\exp \left[ i \frac{\pi}4 \left( \begin{array}{cc}
         \sigma_j & 0 \\
        0 & \sigma_j \end{array} \right) \right] \otimes {\bf 1}_{N_f}.
\label{R}
\end{equation}

If the NNN fermion kernel is taken to be a truncated
SLAC derivative \cite{DWY}, then both couplings $J_1$ and
$J_2$ are positive, and $\hbox{$J_2=J_1/8$}$. If we argue,
however, that the strong-coupling Hamiltonian is derived by
block-spin transformations applied to a short-distance
Hamiltonian, then we cannot say much about the couplings that
appear in it.
We will assume that couplings in the effective
Hamiltonian fall off rapidly with distance; indeed we will assume that
\begin{equation}
\label{strengths}
0<J_2 \ll J_1/N_c.
\end{equation}
This means that we take as our starting point the (globally degenerate)
vacuum determined in \cite{paper1} for the NN theory with $O(1/N_c)$
corrections.
The NNN interaction is a perturbation on this vacuum and its
excitations.

\section{The ground state}
\label{sec:gs} Here we give a short reprise of the results of
\cite{paper1,NNN} for the ground states of the NN and NNN $\sigma$
models with a uniform baryon density $B_\bn=B>0$ (i.e., a uniform
value of $m>N/2$).

The overall factor of $N_c$ in \Eq{S} allows a systematic
treatment in orders of $1/N_c$.  In leading order, the ground
state is found by minimizing the action, which gives field
configurations that are time-independent and that minimize the
interaction. Minimizing the NN interactions results in a locally
degenerate ground state: We assign $\sigma=\Lambda$ on the even
sites and let the $\sigma$ field on each of the odd sites wander
freely in $U(m)/[U(2m-N)\times U(N-m)]$, a submanifold of
$U(N)/[U(m)\times U(N-m)]$. Since the odd sites are independent,
the degeneracy is exponential in the volume.\footnote{This
degeneracy is not removed by the NNN interactions. That is why we
consider the $O(1/N_c)$ corrections first.} In
Ref.~\onlinecite{paper1} we showed that $O(J_1/N_c)$ fluctuations
generate a ferromagnetic interaction among the odd sites, causing
them to align to a common value (``order from disorder''
\cite{odo}).

The resulting ground state has a N\'eel structure. The even sites
break $U(N)$ to $\hbox{$U(m)\times U(N-m)$}$ and then the odd
sites break the symmetry further to $\hbox{$U(2m-N)\times U(N-m)
\times U(N-m)$}$. We can write explicitly
\begin{eqnarray}
\sigma_{\text{even}}&=&U\Lambda_{\text{even}} U^\dag= U\left( \begin{array}{cc}
    \bm1_m & 0   \\
    0 & -\bm1_{N-m}  \end{array} \right)U^\dag,\nonumber\\[2pt]
\sigma_{\text{odd}}&=&U\Lambda_{\text{odd}} U^\dag= U\left( \begin{array}{ccc}
    \bm1_{2m-N} & 0 &0  \\
    0 & -\bm1_{N-m}&0\\
    0&0&\bm1_{N-m}  \end{array} \right)U^\dag.\label{ansatz}
\end{eqnarray}
The matrix $U\in U(N)$ represents the global degeneracy due to
spontaneous symmetry breaking.

We showed in Ref.~\onlinecite{NNN} that the NNN interactions partially remove
this global degeneracy.
We made the ansatz
\begin{equation}
 U=\frac1{\sqrt{2}} \left( \begin{array}{cc}
    u & u   \\
    -u & u  \end{array} \right),
\label{eq:Uo}
\end{equation}
and showed that it yields a ground state.
When
$m \ge 3N/4$, the matrix $u$ is free to take any value within $U(N/2)$, but
a $U(2m-3N/2) \times U(N-m) \times U(N-m)$ subgroup of $U(N/2)$ acts trivially
in \Eq{ansatz} (i.e., matrices in this subgroup give the same field
configuration as choosing $u={\bf 1}$).
Vacua that are associated with different {\em nontrivial\/} choices of $u$ are
in general inequivalent, and give different realizations of the
$U(N_f)\times U(N_f)$ symmetry of the theory.
Since these vacua are not related by symmetry transformations, there is nothing
to prevent lifting of the degeneracy in higher orders in $1/N_c$.
In the sequel, we set $u={\bf 1}_{N/2}$.
This gives the vacuum with the largest symmetry accessible via the ansatz
(\ref{eq:Uo}).

For $m<3N/4$, $u$ was found numerically by minimizing the NNN
energy (\ref{S}). In view of what happens for $m \ge 3N/4$ this
may be only one point in a degenerate manifold of ground
states.\footnote{We have found, in fact, one case where a {\em
different\/} ansatz gives a more symmetric ground state than
\Eq{eq:Uo}. This is the case $(N=12,m=8)$, i.e., $(N_f=3,B=2)$.}
We emphasize that the degeneracy of these vacua is not related to
the global $U(N_f)\times U(N_f)$ chiral symmetry. It is an
accidental global degeneracy of the ground state.

The symmetries of these ground states are summarized in Table
\ref{table1}, reproduced from \cite{NNN}. In general, both chiral
symmetry and discrete lattice rotations are broken; in some cases
a symmetry under rotations around the $z$ axis survives. Note that
if we remove the unphysical axial $U(1)$ symmetry from the
$\sigma$ model, all its realizations will also drop from Table
\ref{table1},  namely, there will be no unbroken axial $U(1)$
symmetries (third column) and no Goldstone bosons corresponding to
a broken axial $U(1)$ (fourth column).

\begin{table}[htb]
\caption{Breaking of $SU(N_f)_L\times SU(N_f)_R\times U(1)_A$ for
all baryon densities (per site) accessible for $N_f\le3$.
\label{table1}}
\begin{ruledtabular}
\begin{tabular}{cccc}

$N_f$   &   $|B|$   &   Unbroken symmetry & Broken charges\\
\hline
    &   0   & $-$  & 1  \\
1   &   1   & $-$  & 1  \\
    &   2   & $U(1)_A$ & 0  \\  \hline
    &   0   & $SU(2)_V$ & 4  \\
    &   1   & $U(1)_{I_3}$ & 6 \\
2   &   2   & $SU(2)_V$ & 4  \\
    &   3   & $U(1)_{I_3}$ & 6  \\
    &   4   & $SU(2)_L\times SU(2)_R\times U(1)_A$ & 0  \\  \hline
    &   0   & $SU(3)_V$ & 9  \\
    &   1   & $U(1)_Y\times SU(2)_V$ & 13\\
    &   2   & $U(1)_Y$ & 16 \\
3   &   3   & $SU(3)_V$ & 9  \\
    &   4   & $U(1)_{I_3}\times U(1)_Y$ & 15 \\
    &   5   & $U(1)_{I_3}\times U(1)_Y\times U(1)_{A'}$ & 14  \\
    &   6   & $SU(3)_L\times SU(3)_R\times U(1)_A$ & 0 \\
\end{tabular}\end{ruledtabular}\end{table}

\section{Spectrum of excitations}
The Goldstone bosons of the NN theory were discussed in \cite{odo}.
As mentioned, they divide into two types.
There are $2(N-m)^2$ bosons of type I with $\omega \sim J_1 |\bk|$ at low
momenta;
these are generalized antiferromagnetic spin waves (and are the only
excitations at zero density).
There are also $2(2m-N)(N-m)$ bosons of type II, that derive their energy
from quantum fluctuations in $O(1/N_c)$.
These are generalized {\em ferromagnetic} magnons
with $\omega \sim (J_1/N_c) |\bk|^2$.

The two types of Goldstone bosons belong to different representations of the
unbroken subgroup $U(2m-N)\times U(N-m) \times U(N-m)$.
This means that they cannot mix to any order in $1/N_c$.
The type I--type II classification is robust in the NN theory.

Now we calculate the effects of the NNN interactions on the
spectrum. In view of \Eq{strengths}, the NN contributions to the
propagators, found in \cite{odo}, remain unchanged. In particular
we can take over the self-consistent determination of the
self-energy of the type II bosons. We need consider the NNN
contributions to the propagators in tree level only. We proceed to
calculate these for $m \ge 3N/4$. In these cases, the calculations
simplify (much as in \cite{NNN}) and we perform them
analytically.\footnote{An exception is $(N=12,m=10)$, where we
have no analytic solution.  See below.} We believe that the
spectra of the other cases have similar features.

In the NN theory the $\sigma$ fields represent fluctuations around the
vacuum (\ref{ansatz}) with $U=1$.
We parametrize them \cite{odo} as
\begin{equation}
\label{even}
\sigma_{\text{even}}=\left( \begin{array}{ccc}
        1-2\chi \chi^{\dag} & -2\chi \pi^{\dag} & -2\chi S
\vspace{0.3cm} \\
        -2\pi \chi^{\dag} & 1-2\pi \pi^{\dag} & -2\pi S \vspace{0.3cm} \\
        -2S \chi^{\dag} & -2S \pi^{\dag} & -1+2\phi^{\dag} \phi \\
\end{array}     \right),
\end{equation}
and
\begin{equation}
\label{odd}
\sigma_{\text{odd}}=\left( \begin{array}{ccc}
       1-2\chi \chi^{\dag} & -2\chi S & 2\chi \pi^{\dag} \vspace{0.3cm} \\
        -2S \chi^{\dag} & -1+2\phi^{\dag} \phi & 2S \pi^{\dag}  \vspace{0.3cm}
\\
        2\pi \chi^{\dag} & 2\pi S & 1-2\pi \pi^{\dag} \\ \end{array} \right).
\end{equation}
Here $\phi$ is an $m\times (N-m)$ complex matrix field written as
\begin{equation}
\phi = \left( \begin{array}{c}
    \chi \\
    \pi \end{array} \right),
\end{equation}
and $S\equiv\sqrt{1-\phi^\dag\phi}$. The field $\pi$ is an
$(N-m)\times(N-m)$ complex matrix, representing the type I
Goldstone bosons. $\chi$ is a $(2m-N)\times (N-m)$ complex matrix
that represents the type II bosons. If $\phi=0$, we have
$\sigma_{\text{even,odd}}=\Lambda_{\text{even,odd}}$, which is the
ground state of the NN theory. We adapt
Eqs.~(\ref{even})--(\ref{odd}) to the NNN theory by rotating them,
\begin{equation}
\label{sig_nnn}
\sigma \to U \sigma U^{\dag},
\end{equation}
with $U$ as given in \Eq{eq:Uo}.
Now $\phi=0$ corresponds to the ground state of the NNN theory.

We substitute Eqs.~(\ref{even})--(\ref{odd}) into the action (\ref{S}).
The rotation $U$ disappears from the kinetic term
and from the NN interaction---they are both $U(N)$ invariant.
This means that the bare
spectra found in \cite{odo}, when $U$ was absent, remain intact.

We write the NNN energy as 
\begin{equation}
\label{Ennn}
E_{\text{nnn}}=\frac{N_cJ_2}4 \sum_{a \bN i} \Tr \alpha_i
\sigma_{a,\bN} \alpha_i \sigma_{a,\bN + 2\ihat},
\end{equation}
where $a=(\text{even, odd})$ and $\bN$ denotes a site on the corresponding fcc sublattice. We rescale $\phi
\rightarrow \phi/\sqrt{N_c}$ and expand \Eq{Ennn} to
second order,
\begin{eqnarray}
E_{\text{nnn}}&=&N_c E_0 +
\frac{J_2}4 \sqrt{N_c} \sum_{a \bN i} \Tr \bar{\alpha}_i \Lambda_a
\bar{\alpha}_i
 \left( \Delta^{(1)}_{a\bN} + \Delta^{(1)}_{a, \bN+2 \ihat} \right) \nonumber \\
               &&+\frac{J_2}4 \sum_{a \bN i} \Tr \bar{\alpha}_i
\Delta^{(1)}_{a\bN} \bar{\alpha}_i \Delta^{(1)}_{a,\bN+2\ihat}
               +\frac{J_2}4 \sum_{a \bN i} \Tr \bar{\alpha}_i
\Lambda_a \bar{\alpha}_i \left( \Delta^{(2)}_{a\bN} +
\Delta^{(2)}_{a,\bN+2\ihat} \right)
\nonumber \\
               &&\quad \qquad +O\left(\frac1{\sqrt{N_c}}\right).
\label{Ennn_1overNc}
\end{eqnarray}
We have defined $\bar{\alpha}_i=U^{\dag} \alpha_i U$, and
$\Delta^{(1,2)}$ correspond to the linear and quadratic deviations
of the $\sigma$ fields from their ground state values. The latter
are given by
\begin{equation} \Delta^{(1)}_{e}=\left(
\begin{array}{ccc}
    0 & 0 & -2\chi \\
    0 & 0 & -2\pi  \\
    -2\chi^{\dag} & -2\pi^{\dag} & 0 \end{array} \right) ,
\qquad  \Delta^{(2)}_{e}=\left( \begin{array}{ccc}
    -2\chi \chi^{\dag} & -2\chi \pi^{\dag} & 0 \\
    -2\pi \chi^{\dag} & -2\pi \pi^{\dag} & 0  \\
    0 & 0 & +2\phi^{\dag} \phi \end{array} \right)
\label{Delta_even}
\end{equation}
on the even sites, and
\begin{equation} \Delta^{(1)}_{o}=V\left(
\begin{array}{ccc}
    0 & 0 & -2\chi \\
    0 & 0 & -2\pi  \\
    -2\chi^{\dag} & -2\pi^{\dag} & 0 \end{array} \right)V^\dag ,
\qquad  \Delta^{(2)}_{o}=V\left( \begin{array}{ccc}
    -2\chi \chi^{\dag} & -2\chi \pi^{\dag} & 0 \\
    -2\pi \chi^{\dag} & -2\pi \pi^{\dag} & 0  \\
    0 & 0 & +2\phi^{\dag} \phi \end{array} \right)V^\dag
\label{Delta_odd}
\end{equation}
on the odd sites.
Here
\begin{equation}
V=\left( \begin{array}{ccc}
    1 & 0 & 0 \\
    0 & 0 & 1 \\
    0 & -1 & 0 \end{array} \right)
\end{equation}
is the matrix that rotates $\Lambda_{\text{even}}$ to
$\Lambda_{\text{odd}}$. It is easy to show that the terms linear
in $\Delta^{(1)}$ vanish.

In view of the block structure of $\Lambda_{\text{even,odd}}$ and
of $U$, as given in Eqs.~(\ref{ansatz})--(\ref{eq:Uo}), and of
$\alpha_i$, it is convenient to decompose $\chi$ for $m\ge3N/4$ as
\begin{equation}
\label{chi}
\chi=\left( \begin{array}{c} \chi_1 \\ \chi_2 \end{array} \right)
\end{equation}
Here $\chi_{1}$ has $N/2$ rows, and $\chi_2$ has $2m-3N/2$ rows.
Both have $N-m$ columns. Substituting into \Eq{Ennn_1overNc} and
omitting the ground state energy we find that the $O(1)$
contribution of the NNN energy depends only on $\chi_1$. The $\pi$
and $\chi_2$ fields do not enter the NNN energy at this order and
remain of type I and of type II, respectively.

We now define a new $N\times N$ matrix $\hat{\chi}$ that contains
only $\chi_1$,
\begin{equation}
\hat{\chi}=\left( \begin{array}{ccc} 0 & 0 & \chi_1 \\
0 & 0 &  0 \\
\chi_1^{\dag} & 0 & 0 \end{array} \right), \label{chi_1}
\end{equation}
and use it to write
\begin{eqnarray}
E_{\text{nnn}}&=& \frac{J_2}4 \sum_{\bN \in \text{fcc}\atop i}
\bigg\lbrace4\Tr\left[ \bar{\alpha}_i \hat{\chi}^e_{\bN}
\bar{\alpha}_i \hat{\chi}^e_{\bN+2\hat{i}}+\bar{\alpha}_{i}
\left(V\hat{\chi}^o_{\bN} V^{\dag}\right)\bar{\alpha}_{i}
\left(V\hat{\chi}^o_{\bN+2\hat{i}} V^{\dag} \right) \right]
\nonumber \\
&& \hskip 2cm
-4\Tr\left[\bar{\alpha}_{i}\Lambda_e\bar{\alpha}_{i}\Lambda_e
\left( \hat{\chi}^e_{\bN} \right)^2 +
\bar{\alpha}_{i}\Lambda_o\bar{\alpha}_{i}\Lambda_o \left(
V\hat{\chi}^o_{\bN} V^{\dag}\right)^2 \right]\bigg\rbrace.
\end{eqnarray}
Next we expand $\hat{\chi}=\chi^\eta\Gamma^\eta$, where
$\chi^{\eta}$ are real and $\Gamma^{\eta}$ are the hermitian
generators of $U(N)$, normalized to
\begin{equation}
\tr \left[ \Gamma^{\eta} \Gamma^{\eta'} \right] = \delta^{\eta \eta'}.
\end{equation}
The form~(\ref{chi_1}) of $\hat\chi$ implies that $\chi^{\eta}\neq
0$ for those generators whose elements
$\left(\Gamma^{\eta}\right)_{\alpha \beta}$ are nonzero for
$\alpha\in [1,N/2]$ and $\beta\in [m+1,N]$ or vice versa. Thus
\begin{equation}
E_{\text{nnn}}= \sum_{a\bN}\sum_{\eta,\eta'} \left[
\chi^{a\eta}_{\bN} \left(N_a\right)^{\eta\eta'}\chi^{a\eta'}_{\bN}
+ \sum_i \chi^{a\eta}_{\bN}
\left(M_{ai}\right)^{\eta\eta'}\chi^{a\eta'}_{\bN+2\hat{i}}\right],
\end{equation}
where
\begin{equation}
\begin{array}{lcl@{\qquad}lcl}
\left(N_e\right)^{\eta \eta'}&=&-J_2\sum_i \Tr \left[
\Gamma^{\eta} \Gamma^{\eta'} \bar{\alpha}_i\Lambda_a
\bar{\alpha}_i \Lambda_a \right],&
\left(M_{ei}\right)^{\eta \eta'}&=&J_2 \Tr \left[ \Gamma^{\eta}
\bar{\alpha}_i \Gamma^{\eta'} \bar{\alpha}_i \right], \\
\left(N_o\right)^{\eta \eta'}&=&-J_2\sum_i \Tr \left[
V \Gamma^{\eta} \Gamma^{\eta'} V^\dag \bar{\alpha}_i\Lambda_a
\bar{\alpha}_i \Lambda_a \right],&
\left(M_{oi}\right)^{\eta \eta'}&=& J_2 \Tr \left[ V \Gamma^{\eta} V^\dag
\bar{\alpha}_i V \Gamma^{\eta'} V^\dag \bar{\alpha}_i \right].
\end{array}
\end{equation}
With a Fourier transform,
\begin{equation}
\chi^{a\eta}_{\bN}=\sqrt{\frac2{N_s\beta}}\sum_{\bk\in
BZ\atop\omega} \chi^{a\eta}_k e^{i\bk \cdot \bN+i\omega \tau},
\label{FT}
\end{equation}
we write the energy in momentum space as
\begin{eqnarray}
E_{\text{nnn}}&=&\sum_{\bk \in BZ \atop \omega >0 }
\chi^{e\dag}_{k}  \left[N_e+N^T_e +
M_e(\bk)+M^\dag_e(\bk)\right]\chi^{e}_{k} \nonumber \\ &&\hskip
1cm+\chi^{o\dag}_{-k} \left[N_o+N^T_o +
M_o(-\bk)+M^\dag_o(-\bk)\right] \chi^{o}_{-k}. \label{SNNNmom}
\end{eqnarray}
Here $M_a(\bk)=\sum_i M_{ai} e^{ik_i}$.

The NN action, including the time derivative and $O(1/N_c)$
self-energy, was written down in  \cite{odo} in terms of the
Fourier transform $\tilde\chi$ of the $(2m-N)\times (N-m)$ matrix
field $\chi$ [\Eq{chi}]:
\begin{eqnarray}
S_{\text{nn}}&=&\sum_{\omega,\bk} \Tr \left[
\left(i\omega-\Sigma_{1,\bk}\right) \tilde\chi^{e\dag}_{k}
\tilde\chi^e_{k}+\left(-i\omega-\Sigma_{1,\bk}\right)
\tilde\chi^{o\dag}_{-k}
\tilde\chi^o_{-k} \right. \nonumber \\
&& \left. \hskip 1.4cm - \Sigma_{2,\bk} \left( \tilde\chi^e_{k}
\tilde\chi^{oT}_{-k}+c.c.\right) \right] .
\end{eqnarray}
The self-energies $\Sigma_a$ are of order $J_1/N_c$ and depend on
$N$ and~$m$.
We set $\chi_2=0$ and repeat the steps leading to \Eq{SNNNmom} to
write $S_{\text{nn}}$ in terms of $\chi_k^{\eta}$,
\begin{equation}
S_{\text{nn}}=\sum_k \chi^{e\dag}_k K_{ee}\chi^{e}_k +
\chi^{o\dag}_{-k} K_{oo} \chi^{o}_{-k} + \chi^{eT}_k
K_{eo}\chi^{o}_{-k}, \label{SNN}
\end{equation}
with the matrices $K_{ee}$, $K_{oo}$, and $K_{eo}$ given by
\begin{eqnarray}
\left( K_{ee} \right)^{\eta\eta'}&=&\frac{i\omega}2 \Tr \left[
\Lambda_e \Gamma^{\eta}\Gamma^{\eta'} \right] -\frac12
\Sigma_{1,\bk} \delta^{\eta\eta'}, \\
\left( K_{oo} \right)^{\eta\eta'}&=&-\frac{i\omega}2 \Tr \left[
\Lambda_e \Gamma^{\eta}\Gamma^{\eta'} \right] -\frac12
\Sigma_{1,\bk} \delta^{\eta\eta'}, \\
\left( K_{eo} \right)^{\eta\eta'}&=&-\Sigma_{2,\bk} \Tr \left[
\Gamma^{\eta}\Gamma^{\eta' T} \right].
\end{eqnarray}
Equation (\ref{SNN}) is to be added to \Eq{SNNNmom} to give the
quadratic action of the type II Goldstone bosons $\chi_1$.

Diagonalizing the quadratic form is straightforward but tedious.
As noted above, only a subset of the generators $\Gamma^\eta$ of $U(N)$
appear in the expansion of \Eq{chi_1}.
Since $\chi_1$ has dimensions $N/2\times(N-m)$, there are
$N(N-m)/2$ pairs of generators in the sum, which we write (similar to
the Pauli matrices $\sigma_x,\sigma_y$) as
$\tilde{\Gamma}_x^{\eta},\tilde{\Gamma}_y^{\eta}$, with $\eta=1,\dots,N(N-m)/2$.
Their coefficients are similarly written as
$\chi_x^{\eta},\chi_y^{\eta}$.
Thus for each $k,\eta$ we have
\begin{equation}
\chi^{\eta}_k=\left( \begin{array}{c}
\chi^{e\eta}_{xk} \\
\chi^{e\eta}_{yk} \\
\chi^{o\eta *}_{x,-k} \\
\chi^{o\eta *}_{y,-k}
\end{array} \right),
\end{equation}
and the action is
\begin{equation}
S=\sum_{\omega>0}\sum_{\bk} \chi^\dag_k \, G^{-1}_k \, \chi_k,
\end{equation}
The inverse propagator is 
\begin{equation}
G^{-1}_k = \left( \begin{array}{cccc}
-\Sigma_{1,\bk} + n_e +m_e  & -\omega   & -\Sigma_{2,\bk}   & 0 \\
\omega   & -\Sigma_{1,\bk} +n_e-m_e & 0 & \Sigma_{2,\bk}  \\
-\Sigma_{2,\bk}   & 0 & -\Sigma_{1,\bk} +n_o+m_o & -\omega   \\
0 & \Sigma_{2,\bk}   & \omega   & -\Sigma_{1,\bk} +n_o-m_o
\end{array} \right) \label{invG},
\end{equation}
where
\begin{eqnarray}
\left( n_a \right)^{\eta\eta'}&=&-2J_2\sum_{i} \tr \left[ (\Gamma_x^{\eta})^2 \bar{\alpha}_i\Lambda_a \bar{\alpha}_i \Lambda_a \right] \delta^{\eta\eta'}, \\
\left( m_a \right)^{\eta\eta'}&=&2J_2 \sum_i \tr \left[ \Gamma_x^{\eta} \bar{\alpha}_i \Gamma_x^{\eta'} \bar{\alpha}_i \right] \cos k_i.
\end{eqnarray}
$\Sigma_{1,2}$ and $\omega$ contain a factor of $\delta_{\eta,\eta'}$.
If we write the $4\times4$ matrix $G^{-1}_k$ in terms of $2\times2$ blocks,
\begin{equation}
G^{-1}_k=\left(
\begin{array}{cc}
A&B\\C&D
\end{array}\right),
\end{equation}
then its determinant is easily calculated via
\begin{eqnarray}
| G^{-1}_k| &=& |C|\,|B-AC^{-1}D|\nonumber\\[3pt]
&=&\left| \begin{array}{cc}
\omega^2+(n_e+m_e-\Sigma_1)(n_o+m_o-\Sigma_1)-\Sigma_2^2 & \omega(n_e+m_e-n_o+m_o) \\
\omega(n_o+m_o-n_e+m_e) & \omega^2+(n_e-m_e-\Sigma_1)(n_o-m_o-\Sigma_1)-\Sigma_2^2 \end{array} \right|.
\nonumber \\ \nonumber \\ \label{detG}
\end{eqnarray}

For $N=4N_f\le12$ (and $m\ge3N/4$), the matrices
$n_e,n_o,m_e,m_o$ all commute, except for the case ($N=12$, $m=10$).  
Dropping this last from consideration, we are left with the values
of $(N_f,B)$ listed in Table \ref{table2}.
For each case, the simultaneous diagonalization of $n_{e,o}$ and $m_{e,o}$
gives the eigenvalues shown.
\begin{table}[htb]
\caption{Simultaneous eigenvalues of $n_{e,o}$ and $m_{e,o}$ (in units of
$2J_2$) for all values of $N_f$ and $B$ considered.
The resulting spectra fall into four classes.
Here $v_z=\cos k_z$ and
$v_{\perp}=2\sin \left(\frac{k_x+k_y}2\right)\sin\left(\frac{k_x-k_y}2\right)$
\label{table2}}
\begin{ruledtabular}
\begin{tabular}{cccccccc}
$N_f$ & $B$ & $n_e$ & $n_o$ & $m_e$ & $m_o$& multiplicity & class \\
\hline
1&1&1&2&$ v_z    $&$ v_\perp$&         &4\\
 & &2&1&$ v_\perp$&$ v_z$    &         &4\\
2&2&1&2&$ v_z    $&$ v_\perp$&$\times3$&4\\
 & &2&1&$ v_\perp$&$ v_z$    &$\times3$&4\\
 & &1&2&$-v_z    $&$-v_\perp$&         &4\\
 & &2&1&$-v_\perp$&$-v_z$    &         &4\\
2&3&0&1&0        &$v_z$      &         &2\\
 & &1&0&$v_z$    &0          &         &2\\
 & &0&2&0        &$v_\perp$  &         &3\\
 & &2&0&$v_\perp$&0          &         &3\\
3&3&1&2&$ v_z    $&$ v_\perp$&$\times6$&4\\
 & &2&1&$ v_\perp$&$ v_z$    &$\times6$&4\\
 & &1&2&$-v_z    $&$-v_\perp$&$\times3$&4\\
 & &2&1&$-v_\perp$&$-v_z$    &$\times3$&4\\
3&5&0&0&0        &0          &$\times2$&1\\
 & &0&1&0        &$v_z$      &         &2\\
 & &1&0&$v_z$    &0          &         &2\\
 & &0&2&0        &$v_\perp$  &         &3\\
 & &2&0&$v_\perp$&0          &         &3\\
\end{tabular}\end{ruledtabular}\end{table}
The zeros
of the determinant~(\ref{detG}) determine the spectrum $\omega(\bk)$, giving
(after $\omega\to i\omega$)
\begin{eqnarray}
\omega^2&=&{\Sigma_1}^2-{\Sigma_2}^2+\frac12\left(n^2_e-m^2_e+n^2_o-m^2_o\right)-\Sigma_1(n_e+n_o) \nonumber \\
&&\pm \Biggl\{\left[ \Sigma_1(n_o-n_e)+\frac12 (n^2_e-m^2_e-n^2_o+m^2_o)\right]^2
\nonumber\\
&&\qquad\qquad+{\Sigma_2}^2\left[(m_e+m_o)^2-(n_e-n_o)^2 \right]\Biggr\}^{1/2}.
\label{poles}
\end{eqnarray}
Because of the symmetries of \Eq{poles}, the spectra of the various cases
shown in Table \ref{table2} fall into four classes.
We examine each class in turn.

\paragraph*{Class 1:}
Here there is no contribution at all from the NNN interaction.  As shown
in \cite{odo}, 
$\Sigma_{1,2}$ are proportional to $J_1/N_c$ and for small momenta they
are quadratic in $|\bk|$; the same holds for the NN
energy $\sqrt{{\Sigma_1}^2-{\Sigma_2}^2}$.
These fields remain isotropic type II Goldstone bosons as in the NN theory.

\paragraph*{Class 2:}
Fields that correspond to the minus sign in \Eq{poles} remain type II, but with
anisotropic dispersion laws of the form
\begin{equation}
\omega^2=c^2\bk^4 \left(1+a\delta \frac{k_z^2}{k^2}\right). \label{case2a}
\end{equation}
The plus sign in \Eq{poles} gives a linear dispersion law, again anisotropic,
\begin{equation}
\label{poles_expand1}
\omega^2=4c_1J_2\left[k_x^2+k_y^2+(1+b\delta)k_z^2\right] \label{case2b}
\end{equation}
The anisotropy in both cases is proportional
to the ratio $\delta \equiv J_2/(12J_1/N_c)$ of NNN to NN couplings.
The coefficients $c$ and $c_1$
are defined as
\begin{eqnarray}
c&=&\left(\frac{d}{d\bk^2}\sqrt{{\Sigma_1}^2-{\Sigma_2}^2}\right)_{\bk=0}, \\
c_1&=&-\left(\frac{d\Sigma_{1}}{d\bk^2}\right)_{\bk=0}>0.
\end{eqnarray}
They are of order $J_1/N_c$.
The coefficients
\begin{equation}
a=\frac{N_c}{6J_1}\frac{c_1^2-c^2}{2c_1c^2}\quad\text{and}\quad
b=\frac{N_c}{6J_1}\frac2{c_1}
\end{equation}
are of order $10^2$ for the cases at hand.

\paragraph*{Classes 3 and 4:}
Taking $\bk=0$ in \Eq{poles} we find that the fields that correspond to the
plus sign get a mass equal to $2J_2$. This is a result of the explicit
breaking of the $U(4N_f)$ symmetry by the NNN interaction terms; these
particles are no longer Goldstone bosons.
The massless bosons in Class 3, corresponding to the minus sign, are
type II bosons described by
\begin{equation}
\label{case3}
\omega^2 = c_1^2\bk^4.
\end{equation}
(This is different from Class 1 where $\omega^2=c^2\bk^4$.)
The massless bosons in Class 4 are again anisotropic, obeying
\Eq{case2b}, and are of type I.

These dispersion relations are correct to $O(\delta)$ for momenta
of $O(\delta)$ or smaller.
In all cases the dispersion relation to $O(1)$ for $\bk^2 \gg \delta$ is
quadratic and
isotropic, unchanged from the NN result presented in \cite{odo}.
Since $\delta$ is a small parameter, this means that in most of the Brillouin
zone the propagator maintains its NN form.
This is the reason why the self consistent calculations in \cite{odo} that
yield $\Sigma_{1,2}$ do not change when we add the NNN
interactions.

\section{Summary and discussion}
\label{summary}

In this work we have studied the non-linear sigma model derived in
\cite{paper1} for the description of lattice QCD with a large density of
baryons.
The model has NN and NNN interactions. 
Building on the results given in \cite{paper1} and \cite{odo} for the
NN theory, and on the study of the NNN ground state presented in \cite{NNN},
we have determined the dispersions relations for the Goldstone bosons in
the NNN theory.

We find that the physics of the NN theory is mostly undisturbed by the NNN interaction.
At leading order, the properties of the type I bosons ($\pi$) and of some of the type II bosons ($\chi_2$) do not change.
The type II bosons grouped in the $\chi_1$ field suffer a variety of fates,
falling into four classes that appear for different values of
$N_f$ and $m$.
Class 1 bosons are unaffected by the NNN perturbation.
Class 2 bosons split into type I and type II, and all 
gain anisotropic contributions of $O(\delta)$ to
their energies.
Some of the Class 3 bosons become massive while other remain unaffected.
In Class 4, some become massive while the others become anisotropic
type I bosons.

The symmetry of the theory, in all cases, is severely broken by the NNN terms---from
$SU(4N_f)$ to $SU(N_f)\times SU(N_f)\times U(1)_A$.
Not surprisingly,
a simple count shows that the total number of massless real fields is far
greater than the number of spontaneously
broken generators of $SU(N_f)\times SU(N_f)\times U(1)_A$,
as shown in Table~\ref{table1}.
The particular NNN interaction we use is simply unable to generate masses
{\em in lowest order\/} for many of the
particles unprotected by Goldstone's Theorem.
This is partly reflected in the accidental
degeneracy of the ground state, which we mentioned below
\Eq{eq:Uo}.
Just as this degeneracy should be lifted in higher orders in $1/N_c$
[beginning with $O(J_2/N_c)$], the corresponding massless excitations
should develop masses.
The only particles protected from mass generation are the minimal number
needed to satisfy Goldstone's theorem (or the Nielsen-Chadha variant).

Another effect that is missing is the mixing of type I and type II Goldstone
bosons, which is certainly permitted when the NNN interaction is turned on.
In \cite{odo} we proved that such a mixing is forbidden in the NN theory, since
the two types of boson belong to different representations of the unbroken
subgroup.
The classification in the NNN theory is less restrictive, and permits mixing
of the bosons. Whether mixing occurs is a dynamical issue that can be
settled only by
calculating to higher order in $1/N_c$. 

To conclude we note that other recent work \cite{Schafer,Sannino} on the high
density regime of QCD---in the continuum---also predicts type II Goldstone bosons and anisotropic dispersion. There, the starting point is 
an effective field theory that describes the low energy dynamics of QCD with nonzero chemical potential $\mu$. For $\mu\neq0$, Lorentz invariance is broken, and the field equations become nonrelativistic. This leads to the emergence of type II Goldstone bosons. In addition, the ground state in \cite{Sannino} can support a nonzero expectation value of vector fields. This breaks rotational
symmetry and makes some of the dispersion relations anisotropic.

\begin{acknowledgments}
We thank Yigal Shamir for helpful
discussions. This work was supported by
the Israel Science
Foundation under grant no.~222/02-1.
\end{acknowledgments}

\end{document}